\documentclass[fleqn,usenatbib]{mnras}

\usepackage{newtxtext,newtxmath}

\usepackage[T1]{fontenc}

\DeclareRobustCommand{\VAN}[3]{#2}
\let\VANthebibliography\thebibliography
\def\thebibliography{\DeclareRobustCommand{\VAN}[3]{##3}\VANthebibliography}

\usepackage{graphicx}	
\usepackage{amsmath}	

\title[Timescale-dependent Optical Variability of Turn-on CLAGNs]
{Timescale-dependent Optical Variability of Turn-on Changing-look Active Galactic Nuclei}

\author[Yu Tao et al.]{%
Yu Tao,$^{1}$
Jie Tang,$^{1}$\thanks{E-mail: tj168@163.com}%
and Xuan Wei$^{1}$\\
$^{1}$School of Physics and Telecommunication Engineering, Shaanxi University of Technology, Hanzhong 723000, China
}

\date{Accepted 2026 August 7. Received 2026 August 5; in original form 2025 December 16}

\pubyear{\the\year{}}

\begin{document}
\label{firstpage}
\pagerange{\pageref{firstpage}--\pageref{lastpage}}
\maketitle

\begin{abstract}
Changing-look active galactic nuclei (CL AGNs) provide a valuable opportunity to study optical variability associated with changes in accretion state. We investigate the optical variability of turn-on CL AGNs, selected through the emergence or strengthening of broad emission lines, using $g$-band light curves from the Zwicky Transient Facility. For a final sample of 106 objects, we measure structure-function amplitudes at three fixed rest-frame timescales, $SF_{30}$, $SF_{150}$ and $SF_{300}$, and examine their dependence on optical luminosity, Eddington ratio, black-hole mass and rest-frame wavelength using Spearman-rank correlations and multivariate regressions. The strongest trends are found on monthly timescales. $SF_{30}$ is significantly anti-correlated with both optical luminosity and Eddington ratio, with Spearman coefficients of $\rho=-0.39$ and $\rho=-0.33$, respectively. These negative trends remain in multivariate regressions after accounting for black-hole mass and rest-frame wavelength. The dependence weakens at longer timescales: $SF_{150}$ shows only weak evidence for luminosity or Eddington-ratio dependence, while $SF_{300}$ shows no robust dependence. Black-hole mass shows no robust single-parameter correlation, and its multivariate coefficients depend on the model form, indicating that its apparent role is affected by covariance with luminosity and Eddington ratio. The rest-frame wavelength term is generally negative, but cannot be separated from redshift in this single-band analysis. These results suggest that monthly optical variability in turn-on CL AGNs is more closely linked to bright-state accretion properties than variability on half-year to year-long timescales.
\end{abstract}

\begin{keywords}
accretion, accretion discs -- galaxies: active -- galaxies: nuclei -- quasars: general -- methods: statistical
\end{keywords}



\section{Introduction}
\label{sec:intro}

Active galactic nuclei (AGNs) are powered by accretion on to central supermassive black holes and radiate across the electromagnetic spectrum. In the optical and ultraviolet bands, the continuum emission is generally associated with thermal radiation from the accretion disc, while the line-emitting gas responds to variations in the ionising continuum over spatially stratified time-scales. AGNs also exhibit stochastic variability on time-scales from days to years, providing a direct probe of the accretion flow and its coupling to the surrounding gas. This variability has commonly been quantified using structure functions, which describe the typical variability amplitude as a function of time lag, although the interpretation of such time-domain measurements can be affected by cadence, baseline length, sampling and other analysis choices \citep[e.g.][]{Ulrich1997,Netzer2015}.

Changing-look AGNs (CL AGNs) are objects that undergo large changes in continuum luminosity and optical spectral type, often accompanied by the appearance or disappearance of broad Balmer emission lines. Such transitions can occur on time-scales of years or less and therefore cannot be fully explained by a static orientation-based unification picture. They are instead widely interpreted as signatures of rapid changes in the accretion flow, although variable obscuration, disc instabilities, changes in disc winds, and variations in the disc--corona structure may also contribute in individual systems \citep[e.g.][]{LaMassa2015,MacLeod2016,Ruan2016,MacLeod2019,RicciTrakhtenbrot2023,Komossa2026}. Recent time-domain surveys and follow-up studies of highly variable AGNs have further shown that large-amplitude optical variability is an efficient way to identify CL AGN candidates and to connect spectral transitions with long-term photometric behaviour \citep[e.g.][]{MacLeod2019,Wang2024}.

Large statistical studies of normal Type-1 AGNs have shown that optical variability amplitude depends on several physical and observational quantities, including rest-frame wavelength, luminosity, black-hole mass, and Eddington ratio. Variability is generally stronger at shorter wavelengths and lower luminosities, while the dependence on black-hole mass and Eddington ratio is more complex and may vary with the time-scale considered \citep[e.g.][]{Schmidt2012,Zuo2012,Sun2014}. However, it remains unclear whether turn-on CL AGNs, as a population selected through the emergence or strengthening of broad emission lines, follow the variability relations established for normal Type-1 AGNs. Since these objects have experienced a major change in continuum and broad-line emission, their optical variability may carry signatures of the physical conditions associated with the turn-on transition. Measuring their variability at fixed rest-frame time-scales therefore offers a way to identify which physical parameters are most closely linked to the variability properties of this population.

In this work, we investigate the optical variability of spectroscopically identified turn-on CL AGNs using $g$-band light curves from the Zwicky Transient Facility (ZTF). The sample is selected as a turn-on CLAGN population, and the ZTF photometry provides the temporal baseline over which the structure functions are measured. We measure structure-function amplitudes at three fixed rest-frame lags: $SF_{30}$, $SF_{150}$, and $SF_{300}$, corresponding approximately to monthly, semiannual, and annual variability. For a final correlation sample of 106 objects with available physical parameters and sufficient light-curve coverage, we examine how these variability amplitudes depend on optical luminosity, Eddington ratio, black-hole mass, and rest-frame wavelength. We use Spearman-rank correlation tests and multivariate regressions to distinguish the primary variability trends from covariances among these parameters.

The paper is organised as follows. Section~\ref{sec:data} describes the sample selection, physical parameters, and ZTF light curves. Section~\ref{sec:methods} presents the structure-function measurements and statistical methods. Section~\ref{sec:results} gives the correlation and multivariate-regression results for $SF_{30}$, $SF_{150}$, and $SF_{300}$. Section~\ref{sec:disc} discusses the implications for the relation between optical variability and accretion-state indicators in turn-on CL AGNs. Finally, Section~\ref{sec:conclusions} summarises the main conclusions.

\section{Sample and data}
\label{sec:data}

We start from the catalogue of 561 changing-look active galactic
nuclei (CL AGNs) presented by \citet{Guo2025}, which was
constructed by comparing repeated SDSS DR16 \citep{Ahumada2020} and DESI DR1 \citep{DESI2026} spectra
and identifying changes in the broad \(\mathrm{H}\beta\),
\(\mathrm{H}\alpha\), or Mg\,\textsc{ii} emission lines. As one of
the largest published spectroscopically selected CL-AGN samples,
with available classifications and consistently derived physical
parameters, this catalogue is well suited to the statistical
variability analysis performed here. In this work, we focus on
turn-on CL AGNs and construct a clean \(\mathrm{H}\beta\)-based
sample for the subsequent analyses. The turn-on classification is adopted from the spectroscopic classifications of \citet{Guo2025}. For these sources, the SDSS spectra represent the faint state, whereas the DESI spectra identify the bright state through the emergence or strengthening of broad emission lines. We do not perform an independent spectral re-classification, but directly adopt the classifications and physical-parameter measurements from the original catalogue.

To ensure that the H$\beta$ emission line is covered by the available spectra and that the relevant spectral measurements are reliable, we restrict the sample to the redshift range $0.2 \le z \le 0.8$. We further remove recurrent changing-look candidates and TDE-like special sources discussed separately by \citet{Guo2025}, and exclude objects for which the SDSS and DESI redshifts differ by more than 0.01. For the core sample, we require reliable DESI bright-state continuum-luminosity measurements with $e_{\log L_{5100}} \le 0.15$ dex. We also require available single-epoch virial black-hole mass estimates based on the DESI bright-state broad H$\beta$ line, with formal uncertainties satisfying $e_{\log M_{\rm BH}} \le 0.30$ dex. The Eddington ratio is calculated using the DESI bright-state $L_{5100}$ and the DESI H$\beta$-based virial black-hole mass. After applying these spectroscopic and physical-parameter cuts, we obtain a core sample of 140 turn-on CL AGNs.

The optical time-domain data are taken from the public Zwicky
Transient Facility (ZTF; \citealt{Bellm2019,Masci2019}) data
releases. For each source in the core sample, we extract the
\(g\)-band light curve using the published coordinates and convert
the observing times to the rest frame using the literature redshift.
We focus on the ZTF \(g\) band because, over
\(0.2\leq z\leq0.8\), it samples rest-frame wavelengths of
approximately \(2620\)--\(3940\,\text{\AA}\), covering the blue to
near-ultraviolet continuum where AGN variability is generally
stronger \citep[e.g.][]{MacLeod2010}. It is therefore well suited to
the short-time-scale continuum variability examined here. Using a
single passband also provides a homogeneous fixed-lag comparison
without introducing band-dependent sampling and photometric
differences. A multi-band analysis would provide a more direct test
of wavelength dependence, but is beyond the scope of the present
work.

The structure-function analysis requires adequate light-curve sampling at the rest-frame time-scales considered in this work. Therefore, not all objects in the 140-object core sample enter the final correlation analysis. After applying the light-curve coverage requirements described in Section~\ref{subsec:sf}, the final correlation sample contains 106 turn-on CL AGNs. For these sources, we measure structure-function amplitudes at three fixed rest-frame lags, $SF_{30}$, $SF_{150}$, and $SF_{300}$, and examine their dependence on optical luminosity, Eddington ratio, black-hole mass, and rest-frame wavelength.

\begin{figure*}
\centering
\includegraphics[width=\textwidth]{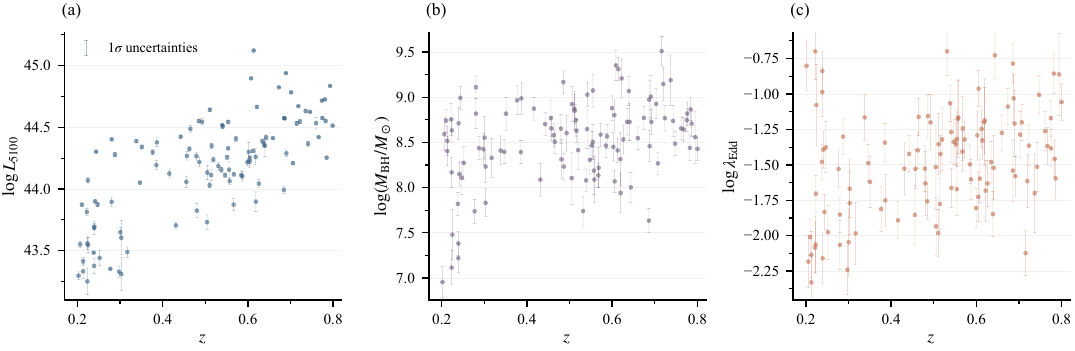}
\caption{Physical-parameter distributions of the final correlation
sample. The three panels show optical continuum luminosity
\(L_{5100}\), black-hole mass \(M_{\rm BH}\), and Eddington ratio
\(\lambda_{\rm Edd}\) as functions of redshift. Error bars show the
\(1\sigma\) measurement uncertainties.}
\label{fig:sample}
\end{figure*}

Figure~\ref{fig:sample} shows the distributions of the main physical parameters of the final correlation sample. The three panels show $L_{5100}$, black-hole mass, and Eddington ratio as functions of redshift, illustrating the coverage of the sample in redshift and physical-parameter space.

\section{Methods}
\label{sec:methods}

\subsection{Structure-function measurements}
\label{subsec:sf}

Structure functions provide an empirical description of the typical
variability amplitude as a function of time lag and have been widely
applied to unevenly sampled astronomical light curves, including
those of AGNs and quasars \citep[e.g.][]{Simonetti1985,VandenBerk2004,deVries2005,Wilhite2008}. They are suitable for the present purpose
because we aim to compare variability amplitudes at predefined
rest-frame lags without requiring each light curve to follow a
specific stochastic-process model. However, structure functions can
be affected by cadence, baseline length, sampling and correlations
among epoch pairs, and apparent features should not automatically
be interpreted as unique physical time-scales
\citep[e.g.][]{Emmanoulopoulos2010,Kankkunen2025}. We therefore use the structure
function as an empirical measure of variability amplitude at three
predefined lags, rather than as a method for identifying a
characteristic time-scale. For each source, we retain photometric measurements with finite observing time, magnitude and magnitude uncertainty, require $0<\sigma_m<0.3$ mag, and use only epochs with ${\tt catflags}=0$ when this flag is available. For light curves with at least 30 clean epochs, we apply a conservative $5\sigma$ median-absolute-deviation clipping to remove strong photometric outliers. All observing times are converted to the rest frame using the redshift of each source.

For every pair of epochs $i$ and $j$, we compute the rest-frame time lag as
\begin{equation}
\Delta t_{\rm rest}=\frac{|t_j-t_i|}{1+z},
\end{equation}
and the magnitude difference as
\begin{equation}
\Delta m_{ij}=m_j-m_i .
\end{equation}
The photometric-noise contribution associated with each pair is
\begin{equation}
\sigma_{ij}^{2}=\sigma_{m,i}^{2}+\sigma_{m,j}^{2}.
\end{equation}
For a set of epoch pairs within a given time-lag bin, the structure-function amplitude is estimated as
\begin{equation}
SF(\Delta t)=
\left[
\frac{\pi}{2}
\left<|\Delta m_{ij}|\right>^{2}
-
\left<\sigma_{ij}^{2}\right>
\right]^{1/2},
\label{eq:sf}
\end{equation}
where the angle brackets denote the mean over all selected epoch pairs. If the quantity inside the square root is non-positive, the measurement is classified as noise dominated and is not used in the correlation or regression analysis. Only positive, noise-corrected structure-function measurements are used when taking logarithms.

We measure the structure function at three fixed rest-frame lags, 30, 150 and 300 d, denoted as $SF_{30}$, $SF_{150}$ and $SF_{300}$, respectively. For each target lag $T$, we select epoch pairs satisfying
\begin{equation}
\left|\log_{10}(\Delta t_{\rm rest})-\log_{10}(T)\right| < 0.125 ,
\end{equation}
corresponding to a total bin width of 0.25 dex in $\log_{10}\Delta t_{\rm rest}$. A valid measurement at a given lag is required to have at least 50 epoch pairs, at least 10 distinct epochs contributing to those pairs, and a rest-frame light-curve baseline longer than the target lag. The uncertainty of each fixed-lag structure-function measurement is estimated using epoch bootstrap resampling. In each bootstrap realization, the light-curve epochs are resampled with replacement, the epoch pairs are reconstructed, and the structure function is remeasured. We use 500 bootstrap realizations and adopt half of the 16th--84th percentile interval as the $1\sigma$ uncertainty.

For clarity, the binned and fixed-lag structure functions serve
different purposes. The binned structure function is evaluated in
a sequence of time-lag bins across the available lag range and is
shown to illustrate the overall lag dependence for the representative
source. The fixed-lag measurements are instead calculated within
the predefined 0.25-dex windows centred on 30, 150, and 300 d for
each individual object. Only the fixed-lag measurements are used in
the subsequent correlation and regression analyses. Because the bin
boundaries and contributing epoch pairs are not necessarily
identical, a fixed-lag measurement need not coincide exactly with
the curve connecting the binned estimates.

\begin{figure*}
\centering
\includegraphics[width=0.82\textwidth]{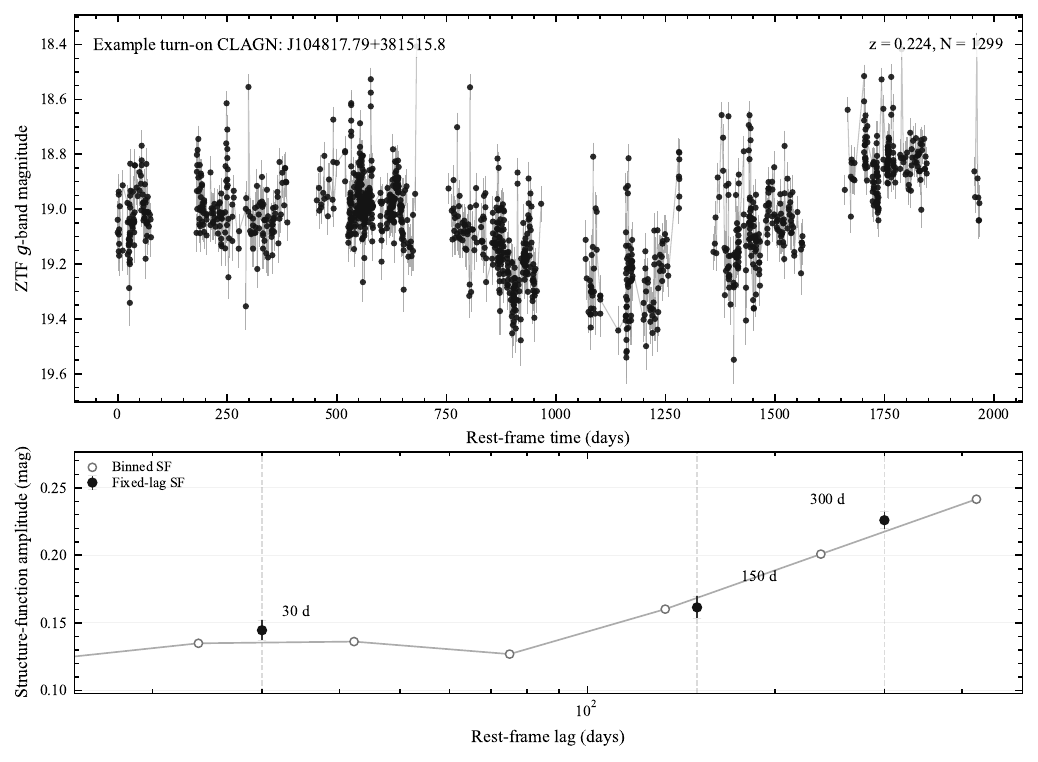}
\caption{Example ZTF $g$-band light curve and structure-function measurements for a representative turn-on CL AGN. The upper panel shows the rest-frame $g$-band light curve of J104817.79+381515.8. The lower panel shows illustrative binned structure-function estimates as open circles, while the filled circles mark the fixed-lag measurements at 30, 150 and 300 d used in the statistical analysis.}
\label{fig:example_lc_sf}
\end{figure*}

Figure~\ref{fig:example_lc_sf} illustrates the structure-function measurement for a representative turn-on CL AGN. The upper panel shows the ZTF $g$-band light curve, while the lower panel shows the corresponding binned structure-function estimates and the three fixed-lag measurements used in the statistical analysis.

Starting from the 140-object core sample, 137 objects have processed
ZTF $g$-band light-curve entries in the structure-function output.
After applying the epoch-pair and coverage requirements, 120, 122,
and 127 measurements could be evaluated at 30, 150, and 300 d,
respectively. Of these, 10, 2, and 0 measurements were classified as
noise dominated, leaving 110, 120, and 127 valid measurements.
Overall, 12 of the 369 evaluated fixed-lag measurements
(3.25 per cent) were noise dominated.

To ensure that the correlations at different time-scales are compared
using the same set of objects, we adopt the common subset with valid
measurements at all three target lags as the final correlation sample.
This yields 106 turn-on CL AGNs. All correlation and regression
analyses below are performed on this common 106-object sample.
Of the 34 objects not included in the final common sample, three have
no matched ZTF light-curve source, three have matched sources but no
available $g$-band measurements, seven have fewer than 30 $g$-band
epochs, nine fail the epoch-pair requirements at one or more target
lags, and 12 have at least one noise-dominated fixed-lag measurement.

\subsection{Correlation and multivariate-regression analyses}
\label{subsec:stat}

We examine the dependence of the structure-function amplitudes on optical luminosity, Eddington ratio, black-hole mass and rest-frame wavelength. The rest-frame wavelength is calculated from the effective wavelength of the ZTF $g$ band, $\lambda_{\rm eff}=4722$~\AA, as
\begin{equation}
\lambda_{\rm rest}=\frac{\lambda_{\rm eff}}{1+z}.
\end{equation}

Because only one observed band is used, \(\lambda_{\rm rest}\) is
a deterministic function of redshift and the two effects cannot be
independently identified in the present analysis. We include
\(\lambda_{\rm rest}\) as an observational wavelength--redshift term
to account for the fact that different sources are sampled at
different rest-frame wavelengths. Its coefficient should therefore
not be interpreted as demonstrating the absence of an intrinsic
redshift dependence.

For the single-parameter correlation analysis, we use Spearman-rank tests to quantify monotonic trends between the structure-function amplitudes and the three main physical parameters, namely optical luminosity, Eddington ratio and black-hole mass. This observational wavelength--redshift term is included as an additional predictor in the multivariate regressions. Since the Spearman coefficient depends only on the rank ordering of the data, using $SF_{\tau}$ or $\log SF_{\tau}$ gives identical correlation coefficients and significance levels. For consistency with the multivariate regressions, we present and discuss the correlation results in terms of $\log SF_{\tau}$, where $\tau=30$, 150 or 300 d.

To assess the stability of the rank-correlation results, we perform
bootstrap resampling and outlier tests. In the bootstrap analysis,
we use 5000 object-level bootstrap realizations, each obtained by
resampling the 106 objects with replacement, and recompute the
Spearman coefficient for each realization. We report the median and
central 95 per cent interval of the bootstrap distribution. For each
structure-function--parameter pair, objects lying below
$Q_1-1.5\,{\rm IQR}$ or above $Q_3+1.5\,{\rm IQR}$ in either
variable are classified as outliers, where
${\rm IQR}=Q_3-Q_1$. After removing these objects, the Spearman
coefficient and $p$-value are recalculated. These tests are used to
distinguish correlations that remain stable under resampling and
outlier removal from marginal trends that are sensitive to sample
fluctuations.

Because luminosity, Eddington ratio, black-hole mass and rest-frame wavelength are not independent of one another, we further perform multivariate regression analyses to separate the primary trends from parameter covariances. Because the structure-function amplitude is positive and empirical relations between AGN variability and physical parameters are commonly examined in logarithmic or approximate power-law form \citep[e.g.][]{VandenBerk2004,Wilhite2008,MacLeod2010,Zuo2012}, we use \(\log SF_{\tau}\) as the response variable in the multivariate regressions. The response variable and all predictors are standardized before fitting, so the regression coefficients describe the relative contribution of each parameter to $\log SF_{\tau}$ after accounting for the other predictors.

We consider two model forms, each fitted separately for $\log SF_{30}$, $\log SF_{150}$ and $\log SF_{300}$. In the luminosity model, the predictors are optical luminosity, black-hole mass and rest-frame wavelength:
\begin{equation}
(\log SF_{\tau})^{\ast}
=
\alpha
+
\beta_{L}(\log L_{5100})^{\ast}
+
\beta_{M}(\log M_{\rm BH})^{\ast}
+
\beta_{\lambda}(\log \lambda_{\rm rest})^{\ast}
+
\epsilon ,
\label{eq:model_lum}
\end{equation}
where $\tau=30$, 150 or 300 d, and the asterisks denote standardized variables. In the Eddington-ratio model, optical luminosity is replaced by Eddington ratio:
\begin{equation}
(\log SF_{\tau})^{\ast}
=
\alpha
+
\beta_{\rm Edd}(\log \lambda_{\rm Edd})^{\ast}
+
\beta_{M}(\log M_{\rm BH})^{\ast}
+
\beta_{\lambda}(\log \lambda_{\rm rest})^{\ast}
+
\epsilon .
\label{eq:model_edd}
\end{equation}

The multivariate regressions are fitted in a Bayesian framework using \textsc{PyMC}. The uncertainty of $\log SF_{\tau}$ is propagated from the bootstrap uncertainty of $SF_{\tau}$ as
\begin{equation}
e_{\log SF_{\tau}}=\frac{e_{SF_{\tau}}}{SF_{\tau}\ln 10},
\end{equation}
and included in the likelihood. For predictors with available measurement uncertainties, we introduce latent true values and model the observed predictors as Gaussian realizations around them. The uncertainty in the rest-frame wavelength is negligible compared with those of the other parameters and is not included explicitly. An intrinsic-scatter term is added in quadrature with the measurement uncertainty of $\log SF_{\tau}$. For each model, we ran four independent Markov chains, each with 3000 tuning steps followed by 5000 retained posterior draws, yielding a total of 20\,000 post-tuning draws. Standard convergence diagnostics indicated satisfactory sampling: all reported global parameters had $\widehat{R}<1.01$, with bulk and tail effective sample sizes exceeding 1600 and 950,
respectively, and no divergent transitions were found. The posterior distributions of the standardized coefficients are used to evaluate the strength and sign of each trend. Corner plots for the six multivariate models are provided in Figures~S1--S6 of the online supplementary material. For each coefficient, we report the posterior mean and the 95 per cent highest-density interval. Coefficients whose intervals exclude zero are generally regarded as statistically supported. The isolated \(M_{\rm BH}\) coefficient in the \(\log SF_{300}\) luminosity model is interpreted as marginal because its lower interval boundary is close to zero and the result is not reproduced in the Eddington-ratio model.

\section{Results}
\label{sec:results}

\subsection{Single-parameter correlations}
\label{subsec:results_corr}

\begin{figure*}
\centering
\includegraphics[width=\textwidth]{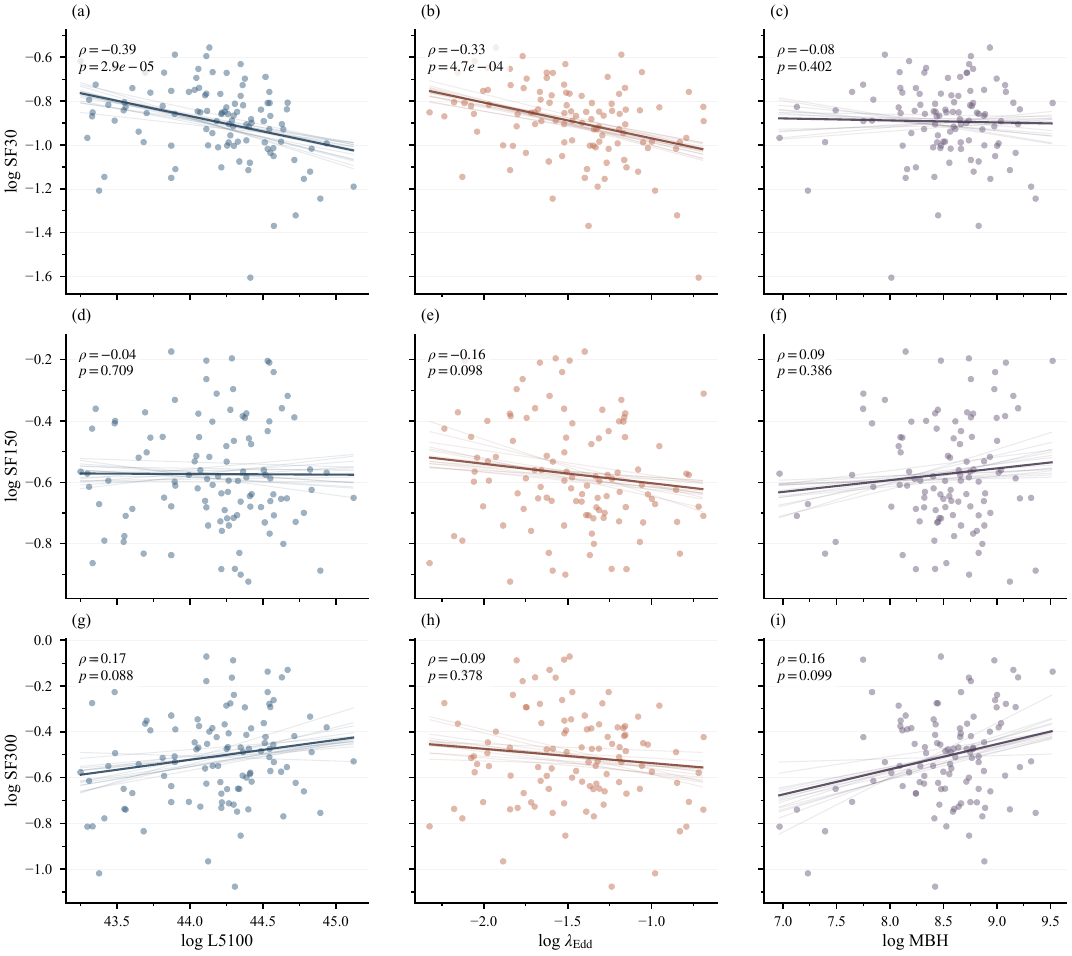}
\caption{Single-parameter relations between the fixed-lag
structure-function amplitudes and physical parameters. Thin lines show ordinary least-squares fits to 15 illustrative object-level bootstrap realizations, while the darker solid lines show the corresponding fit to the original 106-object sample. The Spearman coefficients and \(p\)-values are shown in each panel.}
\label{fig:corr}
\end{figure*}

We first examine the single-parameter relations between the fixed-lag structure-function amplitudes and the main physical parameters. Figure~\ref{fig:corr} shows $\log SF_{30}$, $\log SF_{150}$ and $\log SF_{300}$ as functions of optical luminosity, Eddington ratio and black-hole mass. The corresponding Spearman-rank correlation coefficients, bootstrap intervals and IQR-clipped results are summarized in Table~\ref{tab:spearman}. Overall, the clearest single-parameter trends are found at the shortest time-scale.

At 30 d, $\log SF_{30}$ is significantly anti-correlated with both $L_{5100}$ and $\lambda_{\rm Edd}$. The correlation with $L_{5100}$ has $\rho=-0.394$ and $p=2.92\times10^{-5}$, while the correlation with $\lambda_{\rm Edd}$ has $\rho=-0.334$ and $p=4.67\times10^{-4}$. Both correlations remain stable in the bootstrap resampling and retain the same sign after the IQR outlier test. After removing the IQR-selected outliers, the luminosity correlation remains significant with $\rho_{\rm IQR}=-0.394$ and $p_{\rm IQR}=3.31\times10^{-5}$, and the Eddington-ratio correlation remains significant with $\rho_{\rm IQR}=-0.320$ and $p_{\rm IQR}=8.79\times10^{-4}$. These results indicate that lower-luminosity and lower-Eddington-ratio turn-on CL AGNs tend to show larger optical variability amplitudes on monthly time-scales. By contrast, $\log SF_{30}$ shows no significant single-parameter correlation with black-hole mass.

At 150 d, the single-parameter correlations are substantially weaker. None of the three examined parameters satisfies our stability criteria when the original Spearman test, bootstrap resampling and IQR-clipped test are considered together. The Eddington-ratio trend remains negative, but its significance is only marginal in the original Spearman test and its bootstrap interval includes zero. Thus, the single-parameter analysis provides at most weak evidence for a parameter dependence at the intermediate time-scale.

At 300 d, no robust single-parameter correlation is found. The correlations with luminosity and black-hole mass have only marginal raw $p$-values, and their bootstrap intervals include zero. The IQR-clipped tests also do not support stable trends. The correlation with Eddington ratio is not significant. Therefore, the single-parameter results suggest that the strongest and most stable variability--parameter connections are the anti-correlations of $SF_{30}$ with optical luminosity and Eddington ratio.

\begin{table*}
\centering
\scriptsize
\caption{Single-parameter Spearman-rank correlations and stability tests for the final 106-object sample.}
\label{tab:spearman}
\begin{tabular}{llcccccccc}
\hline
Lag & Parameter & $\rho$ & $p$ & $\rho_{\rm boot}$ & 95 per cent boot. interval & $N_{\rm IQR}$ & $\rho_{\rm IQR}$ & $p_{\rm IQR}$ & Assessment \\
\hline
$\log SF_{30}$  & $\log L_{5100}$          & -0.394 & $2.92\times10^{-5}$ & -0.393 & [-0.557, -0.202] & 105 & -0.394 & $3.31\times10^{-5}$ & Robust \\
$\log SF_{30}$  & $\log \lambda_{\rm Edd}$ & -0.334 & $4.67\times10^{-4}$ & -0.333 & [-0.506, -0.148] & 105 & -0.320 & $8.79\times10^{-4}$ & Robust \\
$\log SF_{30}$  & $\log M_{\rm BH}$        & -0.082 & 0.402 & -0.081 & [-0.278, 0.121] & 100 & -0.149 & 0.138 & Not robust \\
\hline
$\log SF_{150}$ & $\log L_{5100}$          & -0.037 & 0.709 & -0.035 & [-0.228, 0.147] & 101 & -0.052 & 0.604 & Not robust \\
$\log SF_{150}$ & $\log \lambda_{\rm Edd}$ & -0.161 & 0.098 & -0.159 & [-0.349, 0.032] & 101 & -0.127 & 0.206 & Not robust \\
$\log SF_{150}$ & $\log M_{\rm BH}$        &  0.085 & 0.386 &  0.085 & [-0.123, 0.281] &  96 & -0.012 & 0.908 & Not robust \\
\hline
$\log SF_{300}$ & $\log L_{5100}$          &  0.167 & 0.088 &  0.168 & [-0.028, 0.353] & 101 &  0.177 & 0.077 & Not robust \\
$\log SF_{300}$ & $\log \lambda_{\rm Edd}$ & -0.087 & 0.378 & -0.085 & [-0.280, 0.114] & 101 & -0.034 & 0.734 & Not robust \\
$\log SF_{300}$ & $\log M_{\rm BH}$        &  0.161 & 0.099 &  0.160 & [-0.047, 0.358] &  96 &  0.017 & 0.867 & Not robust \\
\hline
\multicolumn{10}{p{0.96\textwidth}}{\footnotesize
Notes. The bootstrap interval is the central 95 per cent interval
of the bootstrap distribution of $\rho$. $N_{\rm IQR}$ is the sample
size after removing objects lying outside the
$Q_1-1.5\,{\rm IQR}$ to $Q_3+1.5\,{\rm IQR}$ range in either
variable for each structure-function--parameter pair. The IQR
columns give the Spearman coefficient and $p$-value after outlier
removal. A relation is classified as robust when the original
correlation has $p<0.05$, the 95 per cent bootstrap interval excludes
zero, and the IQR-clipped correlation retains the same sign with
$p<0.05$.
}
\end{tabular}
\end{table*}

\subsection{Multivariate regression results}
\label{subsec:results_multi}

\begin{figure*}
\centering
\includegraphics[width=\textwidth]{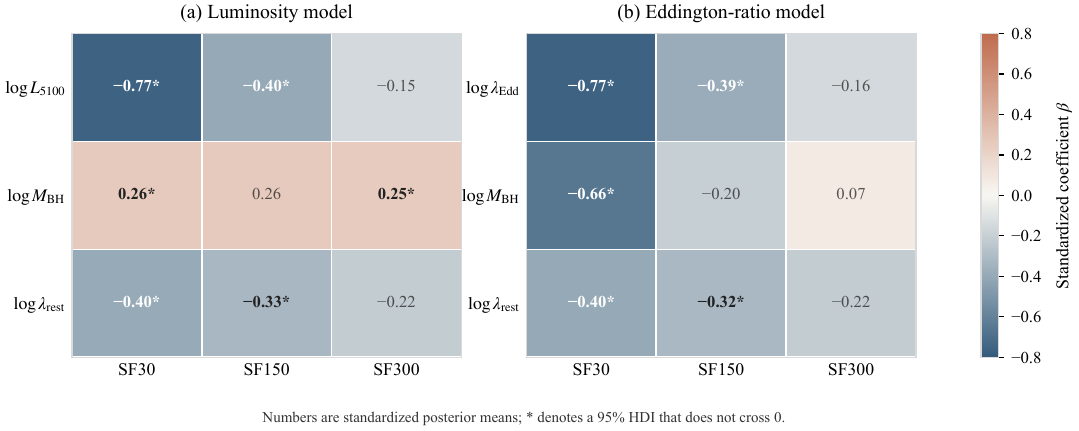}
\caption{Standardized posterior mean coefficients for the luminosity
model (left) and Eddington-ratio model (right).}
\label{fig:multi_reg}
\end{figure*}

The single-parameter correlations do not by themselves separate the effects of luminosity, Eddington ratio, black-hole mass and rest-frame wavelength, because these quantities are correlated with one another in the sample. We therefore use the multivariate Bayesian regressions described in Section~\ref{subsec:stat} to identify which trends remain after accounting for parameter covariances. Figure~\ref{fig:multi_reg} shows the standardized regression coefficients, and Table~\ref{tab:bayes_reg} lists the posterior means and 95 per cent highest-density intervals.

The multivariate regressions confirm that the strongest parameter dependence occurs at 30 d. In the luminosity model, the coefficient of
\(\log L_{5100}\) is negative, with \(\beta=-0.767\) and a 95 per cent HDI of \([-0.998,-0.520]\). Thus, the short-time-scale variability amplitude decreases with increasing optical luminosity after controlling for black-hole mass and rest-frame wavelength. In the Eddington-ratio model, \(\log\lambda_{\rm Edd}\) also has a robust negative coefficient, with \(\beta=-0.766\) and a 95 per cent HDI of \([-0.990,-0.528]\). The coefficient associated with the sampled rest-frame wavelength is also negative in both 30-d models. Its sign is consistent with stronger variability at shorter wavelengths, although the single-band analysis does not separate wavelength from redshift.

The black-hole-mass term behaves differently from the luminosity, Eddington-ratio and wavelength terms. At 30 d, the inferred black-hole-mass coefficient changes sign between the luminosity and Eddington-ratio models. This sign change reflects the covariance among $L_{5100}$, $M_{\rm BH}$ and $\lambda_{\rm Edd}$. We therefore do not interpret black-hole mass as an independent primary driver of the variability amplitude, even when an individual model yields a coefficient whose interval does not include zero.

At 150 d, the multivariate regressions provide weaker evidence for similar trends. The luminosity and Eddington-ratio coefficients remain negative in their respective models, and the rest-frame wavelength term is also negative in both models. However, the coefficient amplitudes are smaller than those at 30 d and the intervals lie close to zero. We therefore interpret the 150-d results as evidence for a weaker intermediate-time-scale dependence rather than as trends as strong as those found at 30 d.

At 300 d, most regression coefficients have 95 per cent HDIs that include zero. Although the black-hole-mass coefficient in the luminosity model formally has a positive interval, its lower bound is very close to zero and the trend is not reproduced in the Eddington-ratio model. We therefore do not regard it as a robust physical result. Overall, the multivariate regressions show no strong evidence that $\log SF_{300}$ depends independently on luminosity, Eddington ratio, black-hole mass or rest-frame wavelength.

Taken together, the single-parameter and multivariate analyses indicate that the variability--parameter connection in turn-on CL AGNs is time-scale dependent. The strongest and most stable trends are found at 30 d, weaker trends are present at 150 d, and no robust dependence is detected at 300 d. The primary robust trends are the negative dependencies of short-time-scale variability on luminosity and Eddington ratio. The coefficient of the observational wavelength--redshift term is also negative at 30 and 150 d, but does not independently distinguish wavelength and redshift effects.

\begin{table*}
\centering
\caption{Standardized coefficients from the multivariate Bayesian regressions.}
\label{tab:bayes_reg}
\begin{tabular}{lllccc}
\hline
Lag & Model & Predictor & $\beta$ & 95 per cent HDI & Assessment \\
\hline
$\log SF_{30}$ & Luminosity & $\log L_{5100}$           
& -0.767 & [-0.998, -0.520] & Robust \\

$\log SF_{30}$ & Luminosity & $\log M_{\rm BH}$         
&  0.258 & [ 0.053,  0.450] & Robust \\

$\log SF_{30}$ & Luminosity & $\log \lambda_{\rm rest}$ 
& -0.396 & [-0.595, -0.199] & Robust \\

$\log SF_{30}$ & Eddington & $\log \lambda_{\rm Edd}$  
& -0.766 & [-0.990, -0.528] & Robust \\

$\log SF_{30}$ & Eddington & $\log M_{\rm BH}$         
& -0.657 & [-0.886, -0.436] & Robust \\

$\log SF_{30}$ & Eddington & $\log \lambda_{\rm rest}$ 
& -0.402 & [-0.591, -0.219] & Robust \\
\hline
$\log SF_{150}$ & Luminosity & $\log L_{5100}$           
& -0.402 & [-0.741, -0.051] & Robust \\

$\log SF_{150}$ & Luminosity & $\log M_{\rm BH}$
& 0.259 & [-0.013, 0.512] & Not robust \\

$\log SF_{150}$ & Luminosity & $\log \lambda_{\rm rest}$ 
& -0.327 & [-0.620, -0.050] & Robust \\

$\log SF_{150}$ & Eddington & $\log \lambda_{\rm Edd}$  
& -0.385 & [-0.704, -0.057] & Robust \\

$\log SF_{150}$ & Eddington & $\log M_{\rm BH}$         
& -0.201 & [-0.517,  0.125] & Not robust \\

$\log SF_{150}$ & Eddington & $\log \lambda_{\rm rest}$ 
& -0.320 & [-0.590, -0.041] & Robust \\
\hline
$\log SF_{300}$ & Luminosity & $\log L_{5100}$           
& -0.155 & [-0.480,  0.159] & Not robust \\

$\log SF_{300}$ & Luminosity & $\log M_{\rm BH}$         
&  0.250 & [ 0.005,  0.496] & Marginal \\

$\log SF_{300}$ & Luminosity & $\log \lambda_{\rm rest}$ 
& -0.223 & [-0.486,  0.052] & Not robust \\

$\log SF_{300}$ & Eddington & $\log \lambda_{\rm Edd}$  
& -0.156 & [-0.449,  0.153] & Not robust \\

$\log SF_{300}$ & Eddington & $\log M_{\rm BH}$         
&  0.071 & [-0.229,  0.367] & Not robust \\

$\log SF_{300}$ & Eddington & $\log \lambda_{\rm rest}$ 
& -0.223 & [-0.484,  0.036] & Not robust \\
\hline
\multicolumn{6}{p{0.92\textwidth}}{\footnotesize
Notes. The coefficients are standardized posterior mean regression
coefficients, and the intervals are 95 per cent highest-density
intervals. Coefficients whose intervals exclude zero are generally
classified as robust. The $M_{\rm BH}$ coefficient in the
$\log SF_{300}$ luminosity model is classified as marginal because
its lower interval boundary is close to zero and the corresponding
coefficient is not supported by the alternative Eddington-ratio
model; it is therefore not treated as a main result.
}
\end{tabular}
\end{table*}

\section{Discussion}
\label{sec:disc}

Our results show that the optical variability of turn-on CL AGNs is linked to physical parameters in a strongly time-scale-dependent way. In the final sample of 106 objects, the clearest trends are found at the shortest rest-frame lag: $SF_{30}$ is significantly anti-correlated with both optical luminosity and Eddington ratio, and these trends remain stable under bootstrap resampling, IQR-based outlier tests and multivariate Bayesian regression. By contrast, the parameter dependence becomes weaker at 150 d and is not robust at 300 d. This suggests that, for turn-on CL AGNs, monthly optical variability is more directly connected to the bright-state accretion properties, whereas half-year to year-scale variability is less well described by single-epoch values of $L_{5100}$, $\lambda_{\rm Edd}$ and $M_{\rm BH}$.

This time-scale dependence is relevant to the physical interpretation of CL AGNs. The defining feature of CL AGNs is not continuum variability alone, but the appearance or disappearance of broad emission lines on relatively short time-scales. Such transitions are commonly discussed in terms of variable obscuration, intrinsic accretion-state changes and the response of the broad-line region to a changing ionizing continuum \citep[e.g.][]{RicciTrakhtenbrot2023,Komossa2026}. For the turn-on objects studied here, the emergence or strengthening of broad Balmer emission lines, together with the availability of bright-state continuum measurements, makes it possible to connect optical variability with bright-state accretion properties in a relatively uniform way. The anti-correlations of $SF_{30}$ with luminosity and Eddington ratio therefore suggest that the short-time-scale optical variability is regulated by the current accretion state.

Previous studies of normal Type 1 AGNs and quasars have shown that
optical variability is generally stronger at lower luminosities,
lower Eddington ratios, and shorter rest-frame wavelengths, although
the reported relations depend on the variability metric and time-scale.
Using ensemble structure functions, \citet{Wilhite2008} measured
the variability strength \(V(\Delta\tau=100\,\mathrm{d})\) for SDSS
quasars. For three luminosity subsamples spanning the same broad
black-hole-mass interval, they found that
\(V(\Delta\tau=100\,\mathrm{d})\) decreased from
\(0.167\pm0.002\) to \(0.142\pm0.002\) and \(0.119\pm0.002\) mag
with increasing luminosity. Using damped-random-walk modelling,
\citet{MacLeod2010} found that the asymptotic variability amplitude
followed
\(SF_{\infty}\propto\lambda_{\rm rest}^{-0.479\pm0.005}\).
For the binned relation between the wavelength-corrected median
\(SF_{\infty}\) and median Eddington ratio, they obtained a power-law
slope of \(-0.23\pm0.03\). Similar luminosity-, Eddington-ratio-,
and wavelength-dependent trends were also reported by
\citet{Zuo2012}.

These measurements are related to, but not identical to, the
fixed-rest-frame-lag structure functions used here. The quantity
\(V(\Delta\tau=100\,\mathrm{d})\) reported by \citet{Wilhite2008}
is an ensemble statistic derived from quasar subsamples, whereas
the DRW-based \(SF_{\infty}\) represents the asymptotic variability
amplitude in the long-time-scale limit. In contrast, our \(SF_{30}\),
\(SF_{150}\), and \(SF_{300}\) values are measured for individual
CL AGNs at selected rest-frame lags. Their absolute amplitudes
should therefore not be compared directly. Instead, we compare the
directions of the reported dependencies on physical parameters.

In our sample, \(SF_{30}\) is anti-correlated with optical luminosity
and Eddington ratio, with Spearman coefficients of \(\rho=-0.394\)
and \(-0.334\), respectively. The luminosity and Eddington-ratio
coefficients remain negative in their respective multivariate models:
the standardized coefficient is \(\beta=-0.767\), with a 95 per cent
HDI of \([-0.998,-0.520]\), in the luminosity model, and
\(\beta=-0.766\), with a 95 per cent HDI of
\([-0.990,-0.528]\), in the Eddington-ratio model. The coefficients
associated with the rest-frame wavelength sampled by the observed
\(g\) band are also negative at 30 d, with \(\beta=-0.396\) and
a 95 per cent HDI of \([-0.595,-0.199]\) in the luminosity model,
and \(\beta=-0.402\) and a 95 per cent HDI of
\([-0.591,-0.219]\) in the Eddington-ratio model.

The luminosity and Eddington-ratio relations are therefore
consistent in sign with those reported for normal broad-line
AGNs. The negative coefficients associated with the sampled
rest-frame wavelength are also consistent with the usual wavelength
dependence of AGN variability. However, because
\(\lambda_{\rm rest}=\lambda_{\rm eff}/(1+z)\) for a single observed
band, rest-frame wavelength and redshift are not independently
identifiable in the present analysis. The wavelength coefficient
therefore cannot by itself distinguish an intrinsic wavelength
dependence from a redshift dependence.

At the same time, turn-on CL AGNs should not be treated simply as normal Type-1 AGNs observed at a single epoch. In normal Type-1 AGNs, optical variability is often interpreted as stochastic fluctuations around a relatively stable long-term accretion state. In CL AGNs, however, the light curves may also contain non-stationary components associated with the changing-look transition itself. This distinction may explain why the physical-parameter dependence is clearest at $SF_{30}$ but becomes much weaker at $SF_{150}$ and is not robust at $SF_{300}$. Monthly variability may mainly trace local disc perturbations, temperature fluctuations or short-time-scale accretion-rate variations in the bright-state accretion flow. In contrast, the half-year and year-scale structure functions may include contributions from transition history, long-term continuum evolution, broad-line-region response and differences in transition phase among individual objects. A single bright-state spectrum can therefore capture part of the short-time-scale variability behaviour, but it is not expected to fully describe the longer-term variability history of a CL AGN.

The comparison with turn-off or off-state CL AGNs is more subtle. The difficulty is not simply that turn-off events are rare. In the DESI--SDSS CLAGN sample of \citet{Guo2025}, the parent catalogue contains comparable numbers of turn-on and turn-off objects. The main difficulty is that the off state is less suitable for homogeneous optical variability--physical-parameter measurements. When the AGN continuum fades, host-galaxy contamination becomes more important. When broad emission lines disappear or become very weak, single-epoch H$\beta$-based estimates of $M_{\rm BH}$ and $\lambda_{\rm Edd}$ can no longer be obtained in the same way as in the bright state. Moreover, off-state optical variability may be affected by broad-line-region fading, variable obscuration, host-galaxy dilution and differences in the stage of the state transition. For this reason, even though turn-off CL AGNs are not numerically negligible, they do not yet provide a directly comparable, homogeneous fixed-lag structure-function benchmark for the off state.

In this sense, the present analysis focuses on the phase in which the relevant physical quantities are most consistently defined. In turn-on CL AGNs, the bright-state continuum and broad H$\beta$ emission allow $L_{5100}$, $M_{\rm BH}$ and $\lambda_{\rm Edd}$ to be measured in a uniform way. The robust anti-correlations of $SF_{30}$ with luminosity and Eddington ratio show that, after the broad lines reappear or strengthen, short-time-scale optical variability remains closely linked to the accretion state. Placed in the broader CLAGN context, this result is consistent with the idea that changing-look transitions occur near a sensitive accretion regime. The bright state reached after a turn-on event may not be identical to a fully settled normal Type-1 phase, but may still retain variability associated with an unsettled accretion state, which is visible in the short-time-scale structure function.

The multivariate regression results further clarify which parameters carry the most direct information. In the luminosity model, the coefficient of $\log L_{5100}$ at 30 d is clearly negative, with $\beta=-0.767$ and a 95 per cent HDI of $[-0.998,-0.520]$. In the Eddington-ratio model, the coefficient of $\log \lambda_{\rm Edd}$ at the same lag is similarly negative, with $\beta=-0.766$ and a 95 per cent HDI of $[-0.990,-0.528]$. These two results are both strong and mutually consistent. At 150 d, the corresponding coefficients remain negative but become weaker, with $\beta=-0.402$ for $\log L_{5100}$ and $\beta=-0.385$ for $\log \lambda_{\rm Edd}$. At 300 d, neither luminosity nor Eddington ratio has a robust coefficient. This progression from strong 30-d trends to weak or absent longer-lag trends provides the main evidence that the variability--accretion-state connection is strongest on monthly time-scales.

By contrast, black-hole mass does not appear to be an independent primary driver of the optical variability amplitude in this sample. In the single-parameter analysis, $M_{\rm BH}$ is not robustly correlated with the structure-function amplitude at any of the three fixed lags. In the multivariate regressions, some black-hole-mass coefficients are formally non-zero, but their sign and strength depend on the model form. In particular, at 30 d the $M_{\rm BH}$ coefficient changes from positive in the luminosity model, $\beta=0.258$ with a 95 per cent HDI of $[0.053,0.450]$, to negative in the Eddington-ratio model, $\beta=-0.657$ with a 95 per cent HDI of $[-0.886,-0.436]$. This sign reversal reflects the covariance among $L_{5100}$, $M_{\rm BH}$ and $\lambda_{\rm Edd}$. The marginal positive $M_{\rm BH}$ coefficient at 300 d in the luminosity model, $\beta=0.250$ with a 95 per cent HDI of $[0.005,0.496]$, is also not reproduced in the Eddington-ratio model. We therefore do not interpret black-hole mass as a stable physical driver of the variability amplitude. The more robust conclusion is that short-time-scale variability is primarily linked to the current radiative output and relative accretion level.

The coefficient associated with the rest-frame wavelength sampled
by the observed ZTF \(g\) band requires a cautious interpretation.
In the multivariate models, this coefficient is negative at 30 and
150 d. At 30 d, it is \(\beta=-0.396\) in the luminosity model and
\(\beta=-0.402\) in the Eddington-ratio model; at 150 d, the
corresponding coefficients are \(\beta=-0.327\) and
\(\beta=-0.320\), respectively. The signs of these coefficients are
consistent with the general tendency for AGN optical variability to
become stronger at shorter rest-frame wavelengths. However, because
\(\lambda_{\rm rest}=\lambda_{\rm eff}/(1+z)\) for a single observed
band, rest-frame wavelength and redshift are not independently
identifiable in the present analysis. We therefore interpret this
predictor as an observational wavelength--redshift term, rather than
as evidence that an intrinsic redshift effect has been excluded.

Before closing, we note several limitations of the present analysis. First, the analysis is based on ZTF \(g\)-band light curves only.
The resulting wavelength--redshift term cannot distinguish an
intrinsic wavelength dependence from a redshift dependence; a
simultaneous multi-band analysis would be required for a direct
test of the wavelength dependence. Secondly, the physical parameters are measured from the DESI bright-state spectra and therefore represent a single spectroscopic state, rather than the full evolution of each object during the changing-look transition. A direct comparison with turn-off CL AGNs, or an analysis restricted to photometrically defined bright-state segments, is left for future work. Finally, the fixed-lag structure functions quantify variability amplitudes at selected time-scales, but do not by themselves identify the detailed temporal phase of the transition in individual sources. These points should be considered when interpreting the weaker longer-time-scale trends.

Taken together, these results place turn-on CL AGNs at the intersection between normal AGN optical variability and changing-look state transitions. On monthly time-scales, they follow the same qualitative behaviour
as normal broad-line AGNs: lower luminosity and lower Eddington
ratio correspond to stronger optical variability, while the negative
observational wavelength--redshift term has a sign consistent with
stronger variability at shorter wavelengths. On longer time-scales, however, the parameter dependence becomes weaker and less stable, indicating that CLAGN light curves cannot be described only as stationary stochastic accretion-disc variability. Fixed-rest-frame-lag structure functions therefore provide a useful empirical way to compare variability amplitudes and their parameter dependence across different rest-frame lags.

\section{Conclusions}
\label{sec:conclusions}

We have investigated the optical variability of spectroscopically identified turn-on CL AGNs using ZTF $g$-band light curves. For a final correlation sample of 106 objects, we measured fixed-lag structure-function amplitudes at 30, 150 and 300 rest-frame days and examined their dependence on optical luminosity, Eddington ratio, black-hole mass and rest-frame wavelength. The main conclusions are as follows.

First, the strongest variability--parameter connection is found at 30 d. The short-time-scale structure-function amplitude $SF_{30}$ is significantly anti-correlated with both optical luminosity and Eddington ratio, and these correlations remain stable under bootstrap resampling and IQR-based outlier tests. The multivariate regressions show that these trends persist after accounting for black-hole mass and rest-frame wavelength.

Second, the dependence on physical parameters weakens with increasing time-scale. At 150 d, the luminosity and Eddington-ratio trends remain negative but are weaker than at 30 d. At 300 d, neither the single-parameter correlations nor the multivariate regressions provide robust evidence for an independent dependence on luminosity or Eddington ratio. This indicates that the variability--accretion-state connection is strongest on monthly rest-frame time-scales.

Third, black-hole mass is not identified as an independent primary driver of the variability amplitude. Although some multivariate coefficients involving $M_{\rm BH}$ are formally non-zero, their sign and robustness depend on the model form. This behaviour is consistent with covariance among luminosity, black-hole mass and Eddington ratio.

Fourth, the coefficient associated with the rest-frame wavelength
sampled by the observed \(g\) band is negative at 30 and 150 d. Its
sign is consistent with stronger variability at shorter wavelengths.
However, because rest-frame wavelength and redshift are
deterministically linked in the single-band data, this result cannot
distinguish an intrinsic wavelength dependence from a redshift
dependence.

Overall, turn-on CL AGNs show short-time-scale optical variability that is closely linked to their bright-state accretion properties. The weaker and less stable trends at longer lags suggest that half-year to year-scale variability is increasingly affected by the non-stationary history of the changing-look transition.

\section*{ACKNOWLEDGEMENTS}

The changing-look AGN classifications and physical-parameter
measurements used in this work are based on the catalogue of
\citet{Guo2025}. This work uses public photometric data from the
Zwicky Transient Facility \citep[ZTF;][]{Bellm2019,Masci2019},
spectroscopic data from SDSS DR16 \citep{Ahumada2020}, and
DESI DR1 \citep{DESI2026}.

The ZTF observations used in this study were obtained with the
Samuel Oschin 48-inch Telescope at Palomar Observatory as part of
the Zwicky Transient Facility project. ZTF is supported by the
National Science Foundation under Grant No.\ AST-1440341 and by
a collaboration involving Caltech, IPAC, the Weizmann Institute
for Science, the Oskar Klein Center at Stockholm University, the
University of Maryland, the University of Washington, Deutsches
Elektronen-Synchrotron, Humboldt University, Los Alamos National
Laboratory, the TANGO Consortium of Taiwan, the University of
Wisconsin--Milwaukee, and Lawrence Berkeley National Laboratory.
ZTF operations are carried out by COO, IPAC, and the University
of Washington. The ZTF data were accessed through the NASA/IPAC
Infrared Science Archive.

Funding for the Sloan Digital Sky Survey IV was provided by the
Alfred P.\ Sloan Foundation, the U.S.\ Department of Energy Office
of Science, and the participating institutions. SDSS also
acknowledges support and computing resources from the Center for
High-Performance Computing at the University of Utah. The SDSS-IV
website is \url{www.sdss4.org}.

SDSS is managed by the Astrophysical Research Consortium on behalf
of the participating institutions of the SDSS Collaboration,
including the Brazilian Participation Group, the Carnegie
Institution for Science, Carnegie Mellon University, the Center for
Astrophysics | Harvard \& Smithsonian, the Chilean Participation
Group, the French Participation Group, Instituto de Astrof\'isica de
Canarias, The Johns Hopkins University, the Kavli Institute for the
Physics and Mathematics of the Universe (IPMU)/University of Tokyo,
the Korean Participation Group, Lawrence Berkeley National
Laboratory, Leibniz Institut f\"ur Astrophysik Potsdam,
Max-Planck-Institut f\"ur Astronomie, Max-Planck-Institut f\"ur
Astrophysik, Max-Planck-Institut f\"ur Extraterrestrische Physik,
National Astronomical Observatories of China, New Mexico State
University, New York University, the University of Notre Dame,
Observat\'orio Nacional/MCTI, The Ohio State University,
Pennsylvania State University, Shanghai Astronomical Observatory,
the United Kingdom Participation Group, Universidad Nacional
Aut\'onoma de M\'exico, the University of Arizona, the University
of Colorado Boulder, the University of Oxford, the University of
Portsmouth, the University of Utah, the University of Virginia, the
University of Washington, the University of Wisconsin, Vanderbilt
University, and Yale University.

This research used data obtained with the Dark Energy Spectroscopic
Instrument (DESI). DESI construction and operations are managed by
Lawrence Berkeley National Laboratory. This work was supported by
the U.S.\ Department of Energy, Office of Science, Office of High
Energy Physics, under Contract No.\ DE--AC02--05CH11231, and by
the National Energy Research Scientific Computing Center, a DOE
Office of Science User Facility operating under the same contract.

Additional support for DESI was provided by the U.S.\ National
Science Foundation, Division of Astronomical Sciences, under
Contract No.\ AST-0950945 to the NSF's National Optical-Infrared
Astronomy Research Laboratory; the Science and Technology
Facilities Council of the United Kingdom; the Gordon and Betty
Moore Foundation; the Heising-Simons Foundation; the French
Alternative Energies and Atomic Energy Commission; the National
Council of Humanities, Science and Technology of Mexico; the
Ministry of Science and Innovation of Spain; and the DESI member
institutions listed at
\url{www.desi.lbl.gov/collaborating-institutions}.

The DESI Collaboration is honoured to conduct scientific research
on I'oligam Du'ag (Kitt Peak), a mountain of particular significance
to the Tohono O'odham Nation. The opinions, findings, conclusions,
and recommendations expressed in this work are those of the authors
and do not necessarily represent the views of the U.S.\ National
Science Foundation, the U.S.\ Department of Energy, or the other
funding agencies listed above.

The analysis in this paper makes use of NumPy
\citep{Harris2020}, SciPy \citep{Virtanen2020}, pandas
\citep{McKinney2010}, Matplotlib \citep{Hunter2007}, PyMC
\citep{AbrilPla2023}, and ArviZ \citep{Kumar2019}.

\section*{Data Availability}

The ZTF light curves used in this work are publicly available from the ZTF data releases via the NASA/IPAC Infrared Science Archive (IRSA). The spectroscopic classifications and physical parameters of the changing-look AGNs are taken from the published SDSS--DESI catalogue of \citet{Guo2025}. The derived fixed-lag structure-function measurements used in the statistical analyses are provided in Appendix~\ref{app:sf_catalogue} and as a machine-readable table in the online supplementary material.



\bibliographystyle{mnras}
\bibliography{refs}




\appendix

\section{Structure-function catalogue}
\label{app:sf_catalogue}

Table~\ref{tab:sf_catalogue_excerpt} shows the first five rows of the structure-function catalogue for the final 106-object correlation sample. The full catalogue is provided as a machine-readable supplementary table. The catalogue includes the source name, coordinates, redshift, SDSS and DESI spectroscopic epochs, bright-state physical parameters with uncertainties, and fixed-lag structure-function measurements at 30, 150 and 300 rest-frame days.

\begin{table*}
\centering
\caption{Example rows from the machine-readable structure-function catalogue.}
\label{tab:sf_catalogue_excerpt}
\scriptsize
\begin{tabular}{lccccccccccc}
\hline
Name & RA & Dec & $z_{\rm DESI}$ & SDSS MJD & DESI MJD & $\log L_{5100}$ & $\log M_{\rm BH}$ & $\log \lambda_{\rm Edd}$ & $SF_{30}$ & $SF_{150}$ & $SF_{300}$ \\
\hline
J001553.74+040039.6 & 3.97392 & 4.01099 & 0.7594 & 55827 & 59498.28 & $44.530\pm0.048$ & $8.656\pm0.128$ & $-1.260\pm0.137$ & $0.097\pm0.012$ & $0.369\pm0.016$ & $0.595\pm0.025$ \\
J001758.76+203407.4 & 4.49483 & 20.56874 & 0.7800 & 56265 & 59557.11 & $44.721\pm0.005$ & $8.443\pm0.138$ & $-0.855\pm0.138$ & $0.048\pm0.024$ & $0.187\pm0.017$ & $0.240\pm0.016$ \\
J002326.09+282112.8 & 5.85873 & 28.35356 & 0.2430 & 57361 & 59503.18 & $44.299\pm0.006$ & $8.993\pm0.125$ & $-1.828\pm0.125$ & $0.100\pm0.004$ & $0.343\pm0.009$ & $0.579\pm0.007$ \\
J003048.87+153438.1 & 7.70362 & 15.57725 & 0.5864 & 56248 & 59551.15 & $44.102\pm0.040$ & $8.567\pm0.133$ & $-1.598\pm0.139$ & $0.169\pm0.039$ & $0.219\pm0.047$ & $0.277\pm0.034$ \\
J004512.26+002827.1 & 11.30111 & 0.47421 & 0.5180 & 55186 & 59571.10 & $44.236\pm0.021$ & $8.419\pm0.220$ & $-1.317\pm0.221$ & $0.143\pm0.012$ & $0.204\pm0.015$ & $0.432\pm0.020$ \\
\hline
\end{tabular}

\begin{minipage}{0.98\textwidth}
\scriptsize
\emph{Notes.} RA and Dec are in degrees. SDSS MJD and DESI MJD give the spectroscopic epochs used to define the faint and bright states. $L_{5100}$ is in erg s$^{-1}$ and $M_{\rm BH}$ is in solar masses. The structure-function amplitudes are in magnitudes. Values in the columns for physical parameters and structure functions are reported as measurement $\pm$ uncertainty. The complete table is available in machine-readable form in the online supplementary material and contains all 106 objects. It also includes $z_{\rm SDSS}$, the SDSS--DESI time separation, logarithmic structure-function amplitudes and their uncertainties, and the numbers of epoch pairs and contributing epochs used for each fixed-lag measurement.
\end{minipage}
\end{table*}


\bsp	
\label{lastpage}
\end{document}